\documentclass[12pt]{article}
\usepackage{
amsfonts,amssymb,epsfig,amsmath}
\usepackage{color}
\renewcommand{\text}[1]{#1}

\newcommand{\be}{\begin{equation}}
\newcommand{\ee}{\end{equation}}
\newcommand{\ben}{\begin{displaymath}}
\newcommand{\een}{\end{displaymath}}
\newcommand{\bea}{\begin{eqnarray}}
\newcommand{\eea}{\end{eqnarray}}
\newcommand{\bean}{\begin{eqnarray*}}
\newcommand{\eean}{\end{eqnarray*}}
\newcommand{\nn}{\nonumber \\}
\newcommand{\ba}{\begin{array}}
\newcommand{\ea}{\end{array}}
\newcommand{\bi}{\begin{itemize}}
\newcommand{\ei}{\end{itemize}}

\newcommand{\reef}[1]{(\ref{#1})}

\def\s{\sigma}

\newcommand{\bbC}{{\mathbb{C}}}

\DeclareMathOperator{\re}{Re}
\DeclareMathOperator{\im}{Im}

\DeclareMathOperator{\vol}{vol}

\newcommand{\uu}{\alpha}
\newcommand{\vv}{\beta}
\newcommand{\ub}{\bar{\alpha}}
\newcommand{\vb}{\bar{\beta}}

\begin{document}

\makeatletter
\renewcommand{\theequation}{\thesection.\arabic{equation}}
\@addtoreset{equation}{section}
\makeatother

\baselineskip 18pt

\begin{titlepage}

\vfill

\begin{flushright}
Imperial/TP/2009/JG/01\\
\end{flushright}

\vfill

\begin{center}
   \baselineskip=16pt
   \begin{Large}\textbf{
        Consistent supersymmetric Kaluza--Klein \\*[5pt] truncations
        with massive modes}
   \end{Large}
   \vskip 1.5cm
    Jerome P. Gauntlett, Seok Kim, Oscar Varela and Daniel Waldram\\
   \vskip .6cm
     \begin{small}
  \textit{Theoretical Physics Group, Blackett Laboratory, \\
        Imperial College, London SW7 2AZ, U.K.}
        \end{small}\\*[.4cm]
        and\\*[.4cm]
      \begin{small}
      \textit{The Institute for Mathematical Sciences, \\
        Imperial College, London SW7 2PE, U.K.}
        \end{small}
   \end{center}

\vfill

\begin{center}
\textbf{Abstract}
\end{center}

\begin{quote}
We construct consistent Kaluza--Klein reductions of $D=11$
supergravity to four dimensions using an arbitrary seven-dimensional
Sasaki--Einstein manifold. At the level of bosonic fields, we extend the known
reduction, which leads to minimal $N=2$ gauged supergravity, to also
include a multiplet of massive fields, containing the breathing mode
of the Sasaki--Einstein space, and still consistent with $N=2$ supersymmetry.
In the context of flux compactifications, the Sasaki--Einstein reductions are
generalizations of type IIA $SU(3)$-structure reductions which include
both metric and form-field flux and lead to a massive universal tensor
multiplet. We carry out a similar analysis for an arbitrary weak $G_2$
manifold leading to an $N=1$ supergravity with massive fields.
The straightforward extension of our results to the case of the
seven-sphere would imply that there is a four-dimensional Lagrangian
with $N=8$ supersymmetry containing both massless and massive spin two
fields. We use our results to construct solutions of M-theory with
non-relativistic conformal symmetry.
\end{quote}

\vfill

\end{titlepage}

\setcounter{equation}{0}



\section{Introduction}

It is now understood that there are very general situations in which
one can perform consistent Kaluza--Klein (KK) reductions of
supergravity theories. Starting with any supersymmetric
solution of $D=10$ or $D=11$ supergravity that is the warped product
of an $AdS_{d+1}$ space with an internal space $M$, it was conjectured
\cite{gv} (see also \cite{Duff:1985jd})
that one can always consistently reduce on the space $M$ to
obtain a gauged supergravity theory in $d+1$ dimensions, incorporating
only the fields of the supermultiplet containing the metric. In the
dual SCFT these fields are dual to the superconformal current
multiplet which includes the energy momentum tensor and $R$ symmetry
currents. This conjecture has now been proven to be true for a
number of general classes of $AdS$ solutions
\cite{Buchel:2006gb}\cite{Gauntlett:2006ai}\cite{gv}\cite{Gauntlett:2007sm}.

One simple class of examples consists of $AdS_5\times SE_5$ solutions of
type IIB supergravity, dual to $N=1$ SCFTs in $d=4$, and $AdS_4\times SE_7$ solutions of $D=11$
supergravity, dual to $N=2$ SCFTs in $d=3$,
where $SE_n$ is an $n$-dimensional Sasaki--Einstein manifold.
In the former case it is known that one can reduce type
IIB supergravity on a $SE_5$ space to get minimal $N=1$ gauged supergravity in
$D=5$ \cite{Buchel:2006gb}.
Similarly, one can reduce $D=11$ supergravity on a $SE_7$ space to get
minimal $N=2$ gauged supergravity in $D=4$ \cite{gv}.
In both cases, the bosonic fields in the lower
dimensional supergravity theory are massless, consisting of the
metric and the gauge field, dual to the energy momentum tensor and the
$R$-symmetry current, respectively. For the special cases when
$SE_5=S^5$ or $SE_7=S^7$ these truncations were shown to be consistent
in \cite{Tsikas:1986rx} and \cite{Duff:1984hn}, respectively. For
these special cases, it is expected or known that there are more
general consistent truncations to the maximal gauged supergravities in
five dimensions (for various partial results see
\cite{Cvetic:1999xp}\cite{Lu:1999bw}\cite{Cvetic:2000nc}\cite{Khavaev:1998fb})
and four dimensions \cite{de Wit:1986iy}, respectively.

Interestingly, it has recently been shown that the consistent KK
reduction of type IIB on a $SE_5$ space of \cite{Buchel:2006gb} can be
generalised to also include some massive bosonic fields
\cite{Maldacena:2008wh}. The bosonic fields included massive gauge
fields as well as massive scalars. One of these massive scalars arises
from the breathing mode of the $SE_5$. Viewing the $SE_5$ space,
locally, as a $U(1)$ fibration over a four-dimensional
K\"ahler-Einstein base, the other massive scalar arises from the mode
that squashes the size of the fibre with respect to the size of the
base. This work thus extends earlier work on including such breathing
and squashing modes for the special case of the five-sphere in
\cite{Bremer:1998zp}\cite{Liu:2000gk}.

In order to understand this in more detail, here we will study similar
extensions of the KK reductions of $D=11$ supergravity on a $SE_7$
space. For the special case of the seven-sphere some results on KK
reductions involving the breathing and squashing modes appear in
\cite{Bremer:1998zp}\cite{Liu:2000gk}. In this paper we shall show
that one can also generalise the KK reduction of \cite{gv} to include
massive fields:
at the level of the bosonic fields we will show that there is a
consistent KK reduction that includes the massless graviton
supermultiplet as well as the massive supermultiplet that contains the
breathing mode. In the off-shell four-dimensional $N=2$ theory, in
addition to the gravity multiplet, the action contains a tensor
multiplet together with a single vector multiplet which acts as a
St\"uckelberg field to give mass to the tensor multiplet. We show that
one can also dualize to get an action containing a massive vector
multiplet with a gauged hypermultiplet acting as the St\"uckelberg
field. This gives a simple example of the mechanism first observed
in~\cite{Louis:2002ny} and then analyzed
in~\cite{TV,Dall,kuzenko,N=1}.

We note that our truncation also has a natural interpretation in terms
of flux compactifications. Viewing the $SE_7$ manifold locally as a
$U(1)$ fibration over a K\"ahler--Einstein manifold, $KE_6$, one can reduce
from M-theory to type IIA. The truncation then has the structure of a
IIA reduction on a six-dimensional $SU(3)$ structure
manifold~\cite{Grana}. The tensor and vector multiplets in
the $N=2$ action correspond to the universal tensor multiplet
which contains the dilaton, the NS two-form $B$ and a complex scalar
arising from a RR potential parallel to the $(3,0)$ form on $KE_6$,
and the universal vector multiplet containing a vector and scalars
that arises from scaling the complexified K\"ahler form. The presence
of the background four-form flux, and the ``metric fluxes'' coming
from the twisting of the $U(1)$ fibration and the fact that the
$(3,0)$ form on $KE_6$ is not closed lead to a gauging of the
four-dimensional theory. This is complementary to the model discussed
in~\cite{Aharony:2008rx} which had a similar structure but considered
different intrinsic torsion in the $SU(3)$ structure. Note that since
our truncation is consistent there are no approximations in analysing
which KK modes should be kept in the four-dimensional theory.

A simple modification of our ansatz leads to an analogous result for a
consistent KK reduction of $D=11$ supergravity on seven-dimensional
manifolds $M_7$ with weak $G_2$ holonomy. Recall that such manifolds
can be used to construct $AdS_4\times M_7$ solutions that are dual to
$N=1$ superconformal field theories in $d=3$. The conjecture of
\cite{gv} is rather trivial for this case since it just says that
there should be a consistent KK reduction to pure $N=1$
supergravity. Here, however, we will see that this can be extended to
include the massive $N=1$ chiral multiplet that contains the breathing
mode of $M_7$. The consistent KK truncation that we construct is
compatible with the general low-energy KK analysis of $D=11$
supergravity reduced on manifolds with weak $G_2$ structure that was
analysed in \cite{House:2004pm}.

Given these results, it is plausible that for $AdS_4\times M_7$
solutions with any amount of supersymmetry $1\le N\le 8$ there is a
consistent KK truncation that includes both the graviton
supermultiplet and the massive breathing mode supermultiplet,
preserving all of the supersymmetry. A particularly interesting
feature for the case of $N=8$ supersymmetry, arising from reduction on
$S^7$, is that the supermultiplet containing the breathing mode now
contains massive spin-2 fields. Thus if our conjecture is correct the
consistent KK reduction would lead to a four-dimensional interacting
theory with both massless and massive spin 2 fields, which has been
widely thought not to exist. We will return to this point in
the discussion section later.

A similar result could also hold for reductions of type IIB on $S^5$
to maximally supersymmetric theories in five spacetime dimensions
containing the massless graviton supermultiplet and the massive
breathing mode supermultiplet, which again contains massive spin 2
fields. What is much more certain, however, is that for reductions on
$SE_5$ one can extend the ansatz of
\cite{Maldacena:2008wh} to be consistent with $N=1$ supersymmetry \cite{gkvw}.

A principal motivation for constructing consistent KK reductions is
that they provide powerful methods to construct explicit solutions.
Starting with the work of \cite{Son:2008ye}\cite{Balasubramanian:2008dm} there has been
some recent interest in constructing solutions of string/M-theory
that possess a non-relativistic conformal symmetry.
In \cite{Maldacena:2008wh} the KK reductions on $SE_5$ spaces were used to construct such
solutions and examples with dynamical exponent $z=4$ and
also $z=2$, and hence possessing an enlarged Schr\"odinger symmetry,
were found. The solutions with $z=2$ were independently found in
\cite{Herzog:2008wg}\cite{Adams:2008wt}.
Here we shall construct similar solutions in $D=11$
supergravity for arbitrary $SE_7$ spaces that exhibit a non-relativistic conformal symmetry
with dynamical exponent $z=3$.

Our presentation will focus on supersymmetric $AdS_4\times
SE_7$ solutions. It is well known that for each supersymmetric
solution there is a ``skew -whiffed'' solution obtained by reversing
the sign of the four-form flux, or equivalently changing the
orientation on  the $SE_7$~\cite{Duff:1984sv}. Apart from the special case of the
round $S^7$ the skew-whiffed solution does not preserve any
supersymmetry, but is known to be perturbatively stable in
supergravity~\cite{Duff:1984sv}. We will show that for the skew-whiffed solutions
there is also a consistent truncation on the $SE_7$ space to the bosonic fields of
a four-dimensional $N=2$ gauged supergravity theory with an $AdS_4$
vacuum that uplifts to the skew-whiffed solution. Our action is a
non-linear extension of one of those considered recently in
\cite{Denef:2009tp} in the context of solutions corresponding to
holographic superconductivity
\cite{Gubser:2008px}\cite{Hartnoll:2008vx}\cite{Hartnoll:2008kx} and
offers the possibility of finding exact embeddings of such solutions
into $D=11$ supergravity.


\section{$D=11$ supergravity reduced on SE$_7$}

Our starting point is the class of supersymmetric $AdS_4\times SE_7$
solutions of $D=11$ supergravity given by
\bea\label{sesol}
  ds^2&=&\tfrac{1}{4}ds^2(AdS_4)+ds^2(SE_7)\nn
  G_4&=&\tfrac{3}{8}\vol(AdS_4)
\eea
where $ds^2(AdS_4)$ is the standard unit-radius metric on $AdS_4$ and
the Sasaki--Einstein metric $ds^2(SE_7)$ is normalised so that the
Ricci tensor is six times the metric (as for a unit-radius round
seven-sphere). The $SE_7$ space has a globally defined one-form $\eta$
that is dual to the Reeb Killing vector, and locally we can write
\be
ds^2(SE_7)=ds^2(KE_6)+\eta\otimes \eta
\ee
where $ds^2(KE_6)$ is a local K\"ahler-Einstein metric with positive
curvature, normalised so that the Ricci tensor is eight times the
metric. On $SE_7$ there is also a globally defined two-form $J$ and a
$(3,0)$-form $\Omega$ that locally define the K\"ahler and complex
structures on $ds^2(KE_6)$ respectively and satisfy $\Omega\wedge
\Omega^*=-8iJ^3/3!$. The Sasaki--Einstein structure implies that
\begin{equation}
\label{SEstructrue}
\begin{aligned}
   d\eta &= 2J \ , \\
   d \Omega &= 4i\eta\wedge \Omega \ . \\
\end{aligned}
\end{equation}
Our conventions for $D=11$ supergravity are as
in~\cite{Gauntlett:2002fz}. For completeness, in
appendix~\ref{app:susy} we show in detail that given these
conventions, together with those for the Sasaki--Einstein structure,
the solution~\eqref{sesol} is indeed supersymmetric.

\subsection{The consistent Kaluza--Klein reduction}
\label{consistentKK}
We now investigate consistent Kaluza--Klein reductions using this
class of solutions. Our ansatz for the metric of $D=11$ supergravity
is given by
\be\label{KKmetT}
ds^2=ds^2_4+e^{2U}ds^2(KE_6)+e^{2V}(\eta+A_1)\otimes(\eta +A_1) \ ,
\ee
where $ds^2_4$ is an arbitrary  metric on a four-dimensional
spacetime, $U$ and $V$ are scalar fields and $A_1$ is a one-form
defined on the four-dimensional space. For the four-form we take
\begin{equation}
\begin{aligned}
   G_4 &= f \textrm{vol}_4
      + H_3 \wedge(\eta+A_1) + H_2 \wedge J
      + H_1 \wedge J \wedge (\eta+A_1) \\ & \qquad \qquad \qquad
      + 2h J \wedge J + {\sqrt 3}\left[
          \chi_1\wedge\Omega +\chi(\eta+A_1)\wedge\Omega
          + \textrm{c.c.} \right] \ ,
\end{aligned}
\end{equation}
where $f$ and $h$ are real scalars, $H_p$, $p=1,2,3$, are real
$p$-forms, $\chi_1$ is a complex one-form, $\chi$ is a complex
scalar on the four-dimensional spacetime and ``$\textrm{c.c.}$'' denotes complex conjugate.

Notice that this ansatz incorporates {\it all} of the constant bosonic
modes that arise from the $G$-structure tensors $(\eta, J,
\Omega)$. It generalises the ansatz considered in \cite{gv}, as we
shall discuss in section 3.1. Together the two scalar fields $U$ and $V$
contain the ``breathing mode'' of the $SE_7$ space and the ``squashing
mode'' that scales the fibre direction with respect to the local $KE_6$
space, as we will discuss more explicitly below. It is also worth
observing that if $\eta,J,\Omega$ instead satisfied
$d\eta=dJ=d\Omega=0$ this ansatz would be the same ansatz that one
would use to reduce $D=11$ supergravity on $S^1\times CY_3$, keeping
the universal $N=2$ vector multiplet, with scalars coming from the
volume mode of the Calabi--Yau, and the universal hypermultiplet.
In particular, we should expect that same off-shell supermultiplet
degrees of freedom to appear in our four-dimensional theory.

We now substitute this ansatz into the equations of motion of $D=11$
supergravity. We will simply summarise the main
results here. More details can be found in
appendix~\ref{app:KK}. By analysing the Bianchi identities
and the equations of motion for the four-form, we find that the
dynamical degrees of freedom turn out to be the four-dimensional
fields $g_{\mu\nu},B_2,B_1,A_1,U,V,h$ and $\chi$ with
\bea\label{pots}
   H_3&=&dB_2\nn
   H_2&=&dB_1+2B_2+hF_2\nn
   F_2&=&dA_1
\eea
Furthermore we find that $H_1=dh$ and $\chi_1=-\frac{i}{4}D\chi$,
where
\be
D\chi\equiv d\chi-4iA_1\chi
\ee
and
\be f=6 e^{-6U-V}(1+h^2+|\chi|^2)\ .
\ee
Note that the expression for $f$ comes from solving \reef{mancity} and
incorporates a convenient integration constant which
fixes the radius of the $AdS$ vacuum and also ensures that the reduced $D=4$ theory
includes the supersymmetric $AdS_4\times SE_7$ solution \reef{sesol}.
The expression for the four-form can be tidied up a little to read
\begin{equation}
\label{KKG4T}
\begin{aligned}
   G_4 &= 6 e^{-6U-V}\left(
           1+h^2+|\chi|^2\right)\vol_4 + H_3 \wedge(\eta+A_1)
      + H_2 \wedge J
      \\ & \qquad
      + dh \wedge J \wedge (\eta+A_1) + 2h J \wedge J
      \\ & \qquad
      + {\sqrt 3}\left[
         \chi(\eta+A_1)\wedge\Omega-\tfrac{i}{4}D\chi\wedge\Omega
         + \textrm{c.c.} \right] \ .
\end{aligned}
\end{equation}

We find that all dependence on the internal $SE_7$ space
drops out of the $D=11$ equations of motion and we are left with
equations of motion for the four-dimensional fields which are written
in appendix~\ref{app:KK}. Thus the ansatz (\ref{KKmetT}), (\ref{KKG4T})
defines a consistent KK truncation. The equations of motion can be
derived from the following four-dimensional action:
\begin{equation}
\label{lagfullchi}
\begin{aligned}
   S &= \int d^4 x \sqrt{-g}e^{6U+V}\Big[
       R + 30(\nabla U)^2+12\nabla U\cdot\nabla V
       - \tfrac{3}{2}e^{-4U-2V}(\nabla h)^2 \\ & \qquad\quad
       - \tfrac32 e^{-6U} |D\chi|^2
       - \tfrac{1}{4}e^{2V}F_{\mu\nu}F^{\mu\nu}
       - \tfrac{1}{12}e^{-2V}H_{\mu\nu\rho}H^{\mu\nu\rho}
       - \tfrac{3}{4}e^{-4U}H_{\mu\nu}H^{\mu\nu} \\ & \qquad\quad
       + 48e^{-2U} - 6e^{-4U+2V}-24h^2 e^{-8U}
       - 18\left(1+h^2 +|\chi|^2 \right)^2e^{-12U-2V}
       \\ & \qquad\quad
       - 24e^{-6U-2V} |\chi|^2 \Big] \\ & \quad
       + \int \Big[
       - 3h H_2\wedge H_2 + 3h^2 H_2\wedge F_2 - h^3 F_2\wedge F_2
       + 6A_1\wedge H_3 \\ & \qquad \quad
       -\tfrac{3i}{4}H_3\wedge (\chi^\ast D\chi-\chi D\chi^\ast)
       \Big] \ .
\end{aligned}
\end{equation}
It is also helpful to write this with respect to the Einstein-frame metric
$g_E\equiv e^{6U+V}g$ and we find
\begin{equation}
\label{lageinfull}
\begin{aligned}
  S &= \int d^4 x\sqrt{-g_E}\Big[
      R_E - 24(\nabla U)^2 -\tfrac32(\nabla V)^2 - 6\nabla U\cdot\nabla V
      \\ & \qquad \quad
      - \tfrac{3}{2}e^{-4U-2V}(\nabla h)^2
      - \tfrac32 e^{-6U} |D\chi|^2
      - \tfrac{1}{4}e^{6U+3V}F_{\mu\nu}F^{\mu\nu}
      \\ & \qquad \quad
      - \tfrac{1}{12}e^{12U} H_{\mu\nu\rho}H^{\mu\nu\rho}
      - \tfrac{3}{4}e^{2U+V}H_{\mu\nu}H^{\mu\nu}
      + 48e^{-8U-V} - 6e^{-10U+V}
      \\ & \qquad \quad
      - 24h^2e^{-14U-V}
      - 18\left( 1+h^2 +|\chi|^2 \right)^2e^{-18U-3V}
      - 24e^{-12U-3V} |\chi|^2
      \Big] \\ & \quad
      + \int \Big[ -3h H_2\wedge H_2
      + 3h^2 H_2\wedge F_2 - h^3 F_2\wedge F_2 + 6A_1\wedge H_3
      \\ & \qquad \quad
      -\tfrac{3i}{4}H_3\wedge (\chi^\ast D\chi-\chi D\chi^\ast)
      \Big] \ .
\end{aligned}
\end{equation}

\subsection{Masses and dual operators}

When we set $H_3=H_2=F=U=V=h=\chi=0$, and thus $f=6$, the equations of
motion are solved by taking the four-dimensional metric to be
$\frac{1}{4}ds^2(AdS_4)$. This ``vacuum solution'' uplifts to give the
$AdS_4\times SE_7$ solution given in \reef{sesol}. We can work out
the masses of the other fields, considered as perturbations about
this vacuum solution, by analysing the quadratic terms in the
Lagrangian \reef{lageinfull}. One immediately deduces that the scalar
fields $h$ and $\chi$ have $m^2_h=40$ and $m^2_\chi=40$.
One can diagonalise the terms involving the
scalar fields $U$ and $V$ by writing
\bea\label{UVdiag}
U&=&-u+\tfrac{1}{3}v\nn
V&=&6u+\tfrac{1}{3}v
\eea
and we find that $m^2_u=16$ and $m^2_v=72$.
Note that in terms of $u$ and $v$ our KK ansatz for the metric \reef{KKmetT}
can be written
\be
ds^2 = e^{-7v/3}ds^2_E + e^{2v/3}\left[e^{-2u}ds^2(KE_6)
   + e^{12u}(\eta+A_1)\otimes(\eta+A_1)\right]
\ee
and we can identify the scalar field $v$ as the ``breathing mode'' and $u$ as the ``squashing mode'' that squashes the
size of the fibre with respect to the size of $KE_6$, preserving
the volume of the $SE_7$ space.

The quadratic action for the fields $A_1,B_1,B_2$ (setting
$U=V=h=\chi=0$) is
\begin{equation}
  \int -\tfrac{1}{2}F_2\wedge *F_2 + \tfrac{1}{2}H_3\wedge *H_3
     -\tfrac{3}{2}(dB_1+2B_2)\wedge*(dB_1+2B_2)
     + 6A_1\wedge H_3\ .
\end{equation}
If one ignores the final term, we see that this has the standard form
for a massless gauge field $A_1$ and a massive two-form $B_2$ with
$B_1$ acting as a St\"uckelberg field. However the presence of the
final term means that the fields are not properly diagonalized.
To find the mass eigenstates, it is helpful to regard
$H_2^\prime\equiv dB_1$ as a basic field by introducing a Lagrange
multiplier one-form $\tilde{B}_1$ and adding a term
\begin{equation}\label{newterm}
  \int 3\tilde{B}_1\wedge dH_2^\prime
\end{equation}
to the action: indeed integrating out $\tilde{B}_1$ brings one back
to the original quadratic action. Integrating out $H_2^\prime$ instead,
we find $H'_2=-*\tilde H_2-2B_2$, where $\tilde H_2\equiv d\tilde B_1$,
and after substitution one obtains the dualised action
\begin{equation}\label{dual}
  \int-\tfrac{1}{2}F_2\wedge *F_2
     + \tfrac{1}{2}H_3\wedge *H_3
     - \tfrac{3}{2}\tilde H_2\wedge*\tilde H_2
     + 6H_3\wedge (\tilde B_1-A_1)\ .
\end{equation}
Continuing we now introduce
\bea
{\cal A}_1&=&\tfrac{1}{2}\big(A_1+3\tilde B_1\big) \ , \nn
{\cal B}_1&=&\tfrac{\sqrt 3}{2}\big(A_1-\tilde B_1\big) \ ,
\eea
so that the action can be written
\begin{equation}\label{dualagain}
  \int-\tfrac{1}{2}d{\cal A}_1\wedge *d {\cal A}_1
  +\tfrac{1}{2}H_3\wedge *H_3
  -\tfrac{1}{2}d{\cal B}_1\wedge *d {\cal B}_1
  -4{\sqrt 3}H_3\wedge {\cal B}_1\ .
\end{equation}
Clearly ${\cal A}_1$ is a massless vector field. The action for the
one-form ${\cal B}_1$ and the two-form $B_2$ appears, for instance,
in~\cite{Minahan:1989vc}. It can be viewed as describing either a
massive vector or a massive two-form field, which are well-known to be
equivalent (see for
example~\cite{Takahashi:1970ev}\cite{Quevedo:1996uu}), with
$m^2=48$. For instance, if one further dualises $\mathcal{B}_1$, one
obtains the standard St\"uckelberg form for a massive two-form.
Alternatively one can dualise the two-form $B_2$ to obtain a
pseudoscalar $a$. This is achieved by adding
\be\label{newterm2}
\int adH_3
\ee
to the action. Integrating out $H_3$, we find that
$H_3=-*(da-4{\sqrt 3}{\cal B}_1)$ and get the action for a massive
vector field $\mathcal{B}_1$
\begin{equation}
 \int -\tfrac{1}{2}d{\cal A}_1\wedge *d {\cal A}_1
    -\tfrac{1}{2}d{\cal B}_1\wedge *d {\cal B}_1
    +\tfrac{1}{2}\big(da-4{\sqrt 3}{\cal B}_1\big)
       \wedge *\big(da-4{\sqrt 3}{\cal B}_1\big) \ .
\end{equation}
In this form, we see that $a$ is a standard  St\"uckelberg scalar
field: using the corresponding gauge symmetry to set $a=0$ reveals
that ${\cal B}_1$ is indeed massive with $m^2=48$.

It is interesting to determine the scaling dimensions of
the operators in the dual SCFT that correspond to the modes we are considering.
The massless vector field, ${\cal A}_1$, has $\Delta=2$
and the massless graviton has $\Delta=3$.
For the scalar fields, using the formula
\be
\Delta=\tfrac{3}{2}\pm\tfrac{1}{2}\sqrt{9+m^2}
\ee
we deduce that the scaling dimensions of $u,h,\chi$ and $v$ are given by
\be
\Delta_u=4,\quad\Delta_h=\Delta_\chi=5,\quad \Delta_v=6
\ee
Finally, for the massive vector field with $m^2=48$, defined by the fields ${\cal B}_1$ and $B_2$,
we can use the formula for a massive $p$-form,
\be
\Delta=\tfrac{3}{2}\pm\tfrac{1}{2}\sqrt{ (3-2p)^2+m^2}
\ee
to deduce that the dual operator has $\Delta=5$.

We will show in the next section that the fields that we have retained
are the bosonic fields of an $N=2$ supergravity theory. In particular
they form the bosonic fields of unitary irreducible
representations of $Osp(2|4)$. The KK modes we have kept are
present for any Sasaki--Einstein seven-manifold and so, in particular,
we can consider the special case of $M(3,2)$ for which the
supermultiplet structure
was analysed in detail in \cite{Fabbri:1999mk}. The massless graviton
and the massless gauge field that we have kept are the bosonic fields
of the massless graviton multiplet, whose field content is summarised
in table 8 of \cite{Fabbri:1999mk}. By analysing the results of
\cite{Fabbri:1999mk} we find\footnote{More specifically, our modes are
  obtained in (3.19) and (3.20) of \cite{Fabbri:1999mk} with
  $M_1=M_2=0$, and hence $J=0$, consistent with the fact that the
  modes are singlets with respect to the $SU(3)$ flavour symmetry of
  this specific Sasaki--Einstein manifold.} that the remaining massive
fields are the bosonic fields of a long vector  multiplet with field
content as in table 3 of \cite{Fabbri:1999mk} with $E_0=4$,
$y_0=0$. Note in particular that with $y_0=0$ the only bosonic modes
with non-zero $R$-charge (``hypercharge'') are the two scalar fields
with $\Delta=5$. These correspond to the $\chi$ fields which indeed
have non-zero $R$-charge since the $(3,0)$-form $\Omega$ in \reef{KKG4T}
carries non-zero $R$-charge.

\subsection{$\mathcal{N}=2$ supersymmetry}

We now show that the Lagrangian~\eqref{lageinfull}
is the bosonic part of an $N=2$ supersymmetric theory. As formulated
it contains, in addition to the $N=2$ supergravity multiplet, a
massive two-form and five scalar fields. The appearance of
supersymmetric theories with a massive two-form in dimensional
reductions with non-trivial fluxes was first observed
in~\cite{Louis:2002ny}. In terms of supermultiplets the two-form and
three scalars should form a tensor multiplet, while the St\"uckelberg
gauge field and the remaining two-scalars form a vector
multiplet.
The general couplings of such $N=2$ theories are discussed
in~\cite{TV,Dall,kuzenko} (see also~\cite{N=1} for the $N=1$
analogue). For the case in hand, it should be possible to dualize to a
massive vector multiplet and a conventional (gauged) hypermultiplet.

As we have noted, our Sasaki--Einstein reduction can also be viewed as
a flux compactification of type IIA supergravity. In particular, if
instead of a reduction on a Sasaki--Einstein manifold
we were considering a reduction on $S^1\times X$ where $X$ is a
Calabi--Yau threefold, the
fields $U$, $\chi$ and the scalar dual of $B_2$ would parameterize a
universal hypermultiplet. Similarly, $2U+V$ and $h$ would be the
scalars for the universal vector multiplet related to rescaling the
metric on the Calabi--Yau space. The kinetic terms of these fields
should be unchanged by going to the Sasaki--Einstein reduction, so our
expectation is that the action~\eqref{lageinfull} can be rewritten as
a gauged universal hypermultiplet coupled to a single universal vector
multiplet. In the following we will show how this structure arises.

Let us first identify the structure before dualizing. The generic
form for the coupling of vector and tensor multiplets has been
discussed in some detail for instance in~\cite{Gunaydin:2005bf}.
We note that
in identifying with the theory of general $N=2$ gauged
supergravities as summarised in appendix B, we should multiply the
overall action in \reef{oact} by a factor of $1/2$, which we will do
in this section only. We can write (half) the
action~\eqref{lageinfull} as
\begin{equation}\label{oact}
  S=\int d^4 x{\sqrt{-g_E}}\left(\tfrac{1}{2}R_E-V\right) + S_V + S_H
\end{equation}
where
\begin{equation}
\begin{aligned}
   S_V &= \frac{1}{2}\int \sqrt{-g_E}\Big[
      -\tfrac{3}{2}(\nabla(2U+V))^2
      -\tfrac{3}{2}\left(e^{-2U-V}\right)^2(\nabla h)^2 \\
      & \qquad\qquad\qquad
         -\tfrac{1}{4}e^{6U+3V}F_{\mu\nu}F^{\mu\nu}
         -\tfrac{3}{4}e^{2U+V}H_{\mu\nu}H^{\mu\nu} \Big] \\
      & \qquad + \frac{1}{2}\int
      -3h H_2\wedge H_2+3h^2 H_2\wedge F_2-h^3 F_2\wedge F_2 \ ,
\end{aligned}
\end{equation}
while
\begin{equation}
\begin{aligned}
  S_H &= \frac{1}{2}\int d^4x{\sqrt{-g_E}}\big[
     -\tfrac{1}{2}\left(\nabla(6U)\right)^2
        - \tfrac{1}{12}e^{12U} H_{\mu\nu\rho}H^{\mu\nu\rho}
        - \tfrac{3}{2}e^{-6U} |D\chi|^2 \big] \\
     & \qquad + \frac{1}{2}\int
     6A_1\wedge H_3
       -\tfrac{3i}{4}H_3\wedge (\chi^\ast D\chi-\chi D\chi^\ast)\ ,
\end{aligned}
\end{equation}
and
\begin{align}
\label{Vdef}
   V &= - 24e^{-8U-V}+3e^{-10U+V}+12h^2e^{-14U-V}
   + 12|\chi|^2e^{-12U-3V} \notag \\ & \qquad
      +9\left(1+h^2+|\chi|^2\right)^2e^{-18U-3V} \ .
\end{align}

If we ignore the $B_2$ term in the definition of $H_2$ in \reef{pots} we see that $S_V$ can be written in
the form of an ungauged vector multiplet action, as summarized in
appendix~\ref{N2susy}, as follows. Introducing $\tau=h+ie^{2U+V}$ we
define $X^I=(1,\tau)$ and the gauge fields
$F^I=\frac{1}{\sqrt{2}}(F_2,-dB_1)$ with $I=0,1$. One then finds the
$\mathcal{N}_{IJ}$ matrix is given by
\begin{equation}
   \mathcal{N}_{IJ}
      = \begin{pmatrix}
            \tau^2(\tau-3h) &\ & 3h\tau \\
            3h\tau &\ & - 3(\tau+h)
        \end{pmatrix}
\end{equation}
together with the corresponding holomorphic prepotential
\begin{equation}
   \mathcal{F}(X) = -\frac{(X^1)^3}{X^0} ,
\end{equation}
giving the K\"ahler potential
\begin{equation}
\label{KVstandard}
   K_V = -\log\left(i\bar{X}^I\mathcal{F}_I
       -iX^I\bar{\mathcal{F}}\right)
    = - \log i(\tau-\bar\tau)^3 \ .
\end{equation}
This is the standard form that arises from flux compactification on a
$SU(3)$ structure manifold~\cite{Louis:2002ny}, with a single K\"ahler modulus.
We also note that, if we ignore the coupling to the vector multiplets, one can dualize the
two-form $B_2$ to get a pseudo-scalar $a$ by adding the term
\reef{newterm2} to $S_H$ giving
$e^{12U}H_3=-*\left[da-\frac{3i}{4}(\chi^*d\chi-\chi
   d\chi^*)\right]$. Then identifying $\rho=4e^{6U}$, $\sigma=4a$ and
$\xi=\sqrt{3}\bar\chi$, we get the standard metric~\eqref{univhyper}
on the universal hypermultiplet space.

To make the full dualization from the massive tensor multiplet
$(e^{6U},B_2,\chi,\bar{\chi})$ to a massive vector multiplet, one must
first dualise the field ${B}_1$ by adding the term
\reef{newterm} to the action and then integrating out $H_2^\prime$. We
now find that
\be
H_2'+2B_2+hF_2=\frac{1}{4h^2+e^{4U+2V}}\left[2h(\tilde H_2+h^2F_2)-e^{2U+V}*(\tilde H_2+h^2F_2)\right]
\ee
where $\tilde{H}_2\equiv d\tilde{B}_1$ as before, and after substitution
one finds a dual action containing gauge fields $A_1,\tilde{B}_1,B_2$.
One can then dualise the two-form $B_2$ to obtain a pseudo-scalar by
adding the term \reef{newterm2}. After integrating out $H_3$ we now find
\be
e^{12U}H_3 = -*\left[da - 6(A_1-\tilde B_1)
   -\tfrac{3i}{4}(\chi^*D\chi-\chi D\chi^*)\right]\ .
\ee

After these dualisations the new expressions for $S_V$ and $S_H$ are
\begin{equation}
\begin{aligned}
   S_V &= \frac{1}{2}\int d^4x{\sqrt{-g_E}}\left[
      -\tfrac{3}{2}\big(\nabla(2U\!+\!V)\big)^2
      -\tfrac{3}{2}\left(e^{-2U-V}\right)^2(\nabla h)^2\right]
      \\ & \quad
      + \frac{1}{2}\int\Big[
         \tfrac{3}{2}\im(\tau+h)^{-1}
         \big( \tilde{H}_2+h^2F_2\big)
         \wedge\ast\big(\tilde{H}_2+h^2F_2\big)
      \\ & \qquad\qquad
      + \tfrac{3}{2}\re(\tau+h)^{-1}
      \big(\tilde{H}_2+h^2F_2\big)\wedge\big(\tilde{H}_2+h^2F_2\big)
      \\ & \qquad\qquad
      -\tfrac{1}{2}\big(e^{2U+V}\big)^3F_2\wedge\ast F_2
      -3h\tilde{H}_2\wedge F_2-h^3 F_2\wedge F_2\Big]
\end{aligned}
\end{equation}
and
\begin{equation}
\begin{aligned}
  S_H &= -\frac{1}{4}\int d^4x{\sqrt{-g_E}}\bigg[
     \big(\nabla(6U)\big)^2
     +3(e^{-6U})\big|d\chi-4iA_1\chi\big|^2
     \\ & \qquad \qquad
     + \left(e^{-6U}\right)^2\left(\nabla a-6(A_1-\tilde{B}_1)
        -\tfrac{3i}{4}(\chi^\ast D\chi-\chi D\chi^\ast)\right)^2
     \bigg] \ .
\end{aligned}
\end{equation}

We now compare $S_V$ with the general gauged $N=2$ action
\reef{N=2action} given in appendix~\ref{N2susy}. If we
identify the gauge fields $\tilde F^I=(F_2,-\tilde H_2)$
and introduce new homogeneous coordinates $\tilde{X}^I=(1,\tau^2)$, we find the
gauge kinetic matrix in~\reef{N=2action} is given by
\begin{equation}\label{coupling}
  \tilde{\mathcal{N}}_{IJ}=\frac{1}{2(\tau+h)}
     \left(\begin{array}{cc}
           -\tau^3\bar\tau & 3h\tau \\
           3h\tau\ & 3
        \end{array}\right)\ ,
\end{equation}
and, since we have dualized the gauge fields, there is a new holomorphic
prepotential
\begin{equation}
   \tilde{\mathcal{F}} = \sqrt{\tilde{X}^0(\tilde{X}^1)^3} \ .
\end{equation}
This indeed correctly reproduces (\ref{coupling}) and leads to the
K\"ahler potential
\begin{equation}
   \tilde{K}_V = - \log i(\tau-\bar{\tau})^3 + \log 2 \ ,
\end{equation}
which agrees with~\eqref{KVstandard} up to a (constant) K\"ahler
transformation.
Notice that $(\tilde X^I, \tilde{\mathcal{F}}_I)$ and
$\tilde{\mathcal{N}}_{IJ}$ can be obtained from $(X^I,\mathcal{F}_I)$
and $\mathcal{N}_{IJ}$ by a symplectic
transformation~\eqref{eq:symplectic} with
\begin{equation}
\label{dualizer}
   \left(\begin{array}{cc} A & B \\ C & D
      \end{array}\right)
   = \sqrt{2}\left(\begin{array}{cc|cc}
         1 &&& \\
         & 0 && -\frac{1}{3} \\
         \hline
         && \frac{1}{2} & \\
         & \frac{3}{2} && 0
      \end{array}\right) ,
\end{equation}
together with a rescaling of the $\tilde{X}^I$ homogeneous coordinates
by $1/\sqrt{2}$. Note that, up to a normalization and as expected given
we are dualizing $B_1$, the matrix~\eqref{dualizer} simply exchanges
the electric and magnetic gauge fields for $F^1$.

Now we consider $S_H$.
Identifying $\rho=4e^{6U}$, $\sigma=4a$, $\xi=\sqrt{3}\bar\chi$
we see that it indeed matches the universal
hypermultiplet form given in~\eqref{univhyper}.
In appendix~\ref{N2susy} we have labelled these coordinates $q^u$,
$u=1,\dots,4$. From the terms
$Dq^u=dq^u-k^u_IA^I$ in \reef{N=2action}
we see that gauging is along Killing vectors
\begin{equation}
\label{nin}
\begin{aligned}
  k_0 &=6\partial_a+4i(\chi\partial_\chi-\bar\chi\partial_{\bar\chi})
  =24\partial_\sigma-4i(\xi\partial_\xi-\bar\xi\partial_{\bar\xi})\ , \\
  k_1 &= 6\partial_a = 24\partial_\sigma\ .
\end{aligned}
\end{equation}
Given the formulae in appendix~\ref{N2susy} for the quaternionic
geometry it is straightforward to calculate that for the Killing vector
$k = \partial_\sigma$ we have the Killing prepotential
\begin{equation}\label{P-sigma}
   P_\sigma = \begin{pmatrix}
           i/4\rho & 0 \\ 0 & -i/4\rho
       \end{pmatrix}
\end{equation}
and for $k=i\xi\partial_\xi-i\bar{\xi}\partial_{\bar{\xi}}$ we have
\begin{equation}\label{P-xi}
   P_\xi = \begin{pmatrix}
           \frac{i}{2}(1-\rho^{-1}\xi\bar{\xi}) &
           -i\xi \rho^{-1/2} \\ -i\bar{\xi}\rho^{-1/2}
           &  -\frac{i}{2}(1-\rho^{-1}\xi\bar{\xi})
           \end{pmatrix} \ .
\end{equation}
The Killing prepotentials $P_I$, corresponding to \reef{nin}, are
therefore
\begin{equation}
  P_0 = 24P_\sigma-4P_\xi \ , \qquad
  P_1 = 24P_\sigma \ .
\end{equation}
Finally substituting these
expressions into the general form~\eqref{N=2V} for the potential $V$
we reproduce~\eqref{Vdef}. This completes our demonstration that our action
is the bosonic action of an $N=2$ supergravity theory

\section{Some further truncations}
\label{other-trunc}

In this section we observe that there are some additional consistent
truncations incorporated in our KK ansatz (\ref{KKmetT}),
(\ref{KKG4T}), compatible with the general equations of motion
contained in appendix~\ref{app:KK}. We begin by observing that it is
consistent to set the complex scalar field $\chi=0$. This is not
surprising as this is the only field in the ansatz that carries
non-zero $R$-charge. The resulting equations of motion can be obtained
from an action obtained by setting $\chi=0$ in \reef{lageinfull}.

\subsection{Minimal gauged supergravity}

It is also consistent to set $U=V=h=\chi=H_3=0$, $f=6$ and
$H_2=-*F_2$. This sets all of the massive fields to zero and we then
find that the equations of motion come from a Lagrangian given by
\be
S=\int d^4x\sqrt{-g}\left[R-F_{\mu\nu}F^{\mu\nu}+24\right]
\ee
This is the
consistent KK reduction on a $SE_7$ to the massless fields of $N=2$ $D=4$ gauged
supergravity that was discussed in \cite{gv}.

It is interesting to ask whether
this truncation to minimal gauged supergravity can be extended to
just include the breathing mode scalar $v$. However, if we take
$h=\chi=H_3=0$, $H_2=-*F_2$ with $U=V=v/3$ and $f=6e^{-7v/3}$ we find
that consistency requires $v=0$ in addition.

\subsection{Scalars}

We next observe that it is possible to consistently truncate to just
the scalar fields plus the metric by setting $H_3=H_2=F_2=0$ and
$\im\chi=0$. The resulting equations of motion follow from
an action which can be obtained by substituting this truncation
directly into the general action \reef{lageinfull}:
\begin{equation}
\label{truncscalar}
\begin{aligned}
   S &= \int d^4x\sqrt{-g_E}\Big[ R_E
        - 42 (\nabla u)^2 -\tfrac72(\nabla v)^2
        - \tfrac{3}{2}e^{-8u-2v}(\nabla h)^2
        - \tfrac32 e^{6u-2v} (\nabla \chi_{\textrm{R}})^2
        \\ & \qquad \qquad
        + 48e^{2u-3v}-6e^{16u-3v}
        - 24h^2 e^{8u-5v}
        - 18(1+h^2 +\chi_{\textrm{R}}^2 )^2e^{-7v}
        \\ & \qquad \qquad
        -24e^{-6u-5v} \chi_{\textrm{R}}^2 \Big]
\end{aligned}
\end{equation}
and we have switched from $U$ and $V$ to $u$ and $v$ via \reef{UVdiag}
and $\chi_{\textrm{R}}=\re\chi$.

In fact, it is also consistent to further set $\chi_{\textrm{R}}=0$ or
$h=0$, or both, and the equations of motion are those that are
obtained by substituting into the action \reef{truncscalar}. Note that for the
case of the seven-sphere setting $\chi_{\textrm{R}}=h=0$, which just
maintains the breathing and squashing mode scalars, was also
considered\footnote{Note that in section 2.2 of \cite{Bremer:1998zp}
  they also consider the truncation with, in the language of this
  paper, $h=\chi=H_2=0$. However, this is not a consistent truncation:
  equation \reef{Bianchi2} implies that $H_3=0$ and then \reef{geom1}
  implies that $F_2=0$.} in \cite{Bremer:1998zp}. Indeed if we
substitute $\chi_{\textrm{R}}=h=0$, $v=-\tilde\varphi/{\sqrt 7}$ and
$u=\tilde\phi/2{\sqrt 21}$ into \reef{truncscalar} we obtain results
equivalent to (2.20) and (2.21) of \cite{Bremer:1998zp}.

A different, further consistent truncation is achieved by setting
$u=0$ and $\chi_{\textrm{R}}=\frac{2}{\sqrt 3}h$ in \reef{truncscalar}. This
truncation generically  breaks supersymmetry down to $N=1$, as we will
see in the next subsection.

\subsection{The weak $G_2$ case}

As we have just noted, it is consistent to set
$H_3=H_2=F_2=0$, $u=0$ (or, equivalently, $U=V$) and $\chi=\frac{2}{\sqrt 3}h$. The resulting equations of motion can be obtained from
an action which can be obtained from (\ref{truncscalar}) and reads:
\begin{equation}
\label{wg2}
\begin{aligned}
   S &= \int d^4 x \sqrt{-g_E}\Big[ R_E
         - \tfrac{7}{2}(\nabla v)^2
         - \tfrac{7}{2}e^{-2v}(\nabla h)^2
         \\ & \qquad \qquad
         + 42e^{-3v} - 56e^{-5v}h^2
         -2\left(3+7h^2\right)^2e^{-7v} \Big] \; .
\end{aligned}
\end{equation}
Note that expanding about the $AdS_4$ vacuum we find $m^2_v=72$,
$m^2_h=40$ and hence $\Delta_v=6$, $\Delta_h=5$.

It is interesting to observe that for this truncation, the KK ansatz
\reef{KKmetT}, \reef{KKG4T} for the $D=11$ fields
can be written
\bea\label{wg2a}
ds^2&=&ds^2_4+e^{2v/3}ds^2(SE_7)\nn
G_4 &=& f \vol_4 +
dh \wedge \varphi
+4h \ast_7\varphi
\eea
where $f=2e^{-7v/3}(3+7h^2)$ and we have introduced the quantities
\bea
\varphi&=&J \wedge \eta + \im\Omega\nn
\ast_7\varphi&=&\tfrac{1}{2}J \wedge J+\eta\wedge \re\Omega
\eea
that satisfy
\be\label{dip}
d\varphi=4 \ast_7\varphi
\ee
Interpreting $\varphi$ as a $G_2$ structure on the seven-dimensional
space $SE_7$, the condition~\eqref{dip} is equivalent to weak $G_2$ holonomy
(i.e. that the cone over the space has $Spin(7)$ holonomy). One can
then generalize by replacing $SE_7$ with an arbitrary space $M$ with weak
$G_2$ holonomy and the ansatz~\reef{wg2a} still gives a consistent
truncation. One would expect such a truncation to have $N=1$
supersymmetry, with the metric lying in an $N=1$ supermultiplet and
the breathing mode in a massive $N=1$ chiral multiplet. In fact,
introducing a complex scalar $\phi = e^{v} + ih$, the $N=1$
supersymmetry of the action~\eqref{wg2} can be explicitly exhibited by
rewriting it in terms of a K\"ahler metric $g_{\phi \bar
  \phi}=\partial_\phi\partial_{\bar\phi} K$ with K\"ahler potential $K
= -7 \log (\phi + \bar \phi)$, and
a superpotential $W = 4\sqrt{2}(7\phi^2 -3)$, as
\be
 S=\int d^4 x
  \sqrt{-g_E}\left[R_E
     - 2g_{\phi \bar \phi} \partial_\mu \phi \partial^\mu \bar \phi
     - 2e^{K} \left(g^{\phi \bar \phi}|D_\phi W|^2 -3 |W|^2 \right)
     \right]
\ee
where $D_\phi W = \partial_\phi W + (\partial_\phi K)W$.
It is worthwhile noting that starting with
the KK ansatz \reef{wg2a} this superpotential can be derived
from the general expression for the form of the superpotential in KK
reductions on manifolds with $G_2$ structure that was obtained in
\cite{House:2004pm}. In contrast to \cite{House:2004pm}, here we have
also shown that this particular KK reduction is a consistent KK
truncation.

We could go one step further and also set $h=0$. We then get the
consistent KK truncation that is valid for any Einstein
seven-manifold, where one keeps only the metric and the breathing mode
scalar. The action is given by
\begin{equation}
\label{gravity+breathing}
  S = \int d^4 x
  \sqrt{-g_E}\left[R_E-\tfrac{7}{2}(\nabla v)^2
  +42e^{-3v}
  -18e^{-7v}\right]
\end{equation}
and we see that $m^2_v=72$ and hence $\Delta_u=6$. The action
(\ref{gravity+breathing}) was first obtained in \cite{Bremer:1998zp}
in the context of the seven-sphere. Specifically
(\ref{gravity+breathing}) can be obtained from (2.6), (2.7) of
\cite{Bremer:1998zp} by setting $\phi=- \sqrt{7} v$, $c=6$, $R_7=42$.

\subsection{Massive vector}

We can also consider a truncation to a metric and a
massive vector field. We now set $U=V=h=\chi=0$,
$f=6$,$H_2=\frac{1}{3}*F_2$ and $H_3=8*A_1$. We now find that {\it
  provided
we restrict to configurations that satisfy}
\be\label{consmav}
F_2\wedge F_2=F_2\wedge *F_2=A_1\wedge *A_1=0\ ,
\ee
the equations of motion can be written
\begin{equation}
\label{eommav}
\begin{aligned}
   d*F_2 &= 48*A_1 \ , \\
   R_{\mu\nu} &= -12g_{\mu\nu} + \tfrac{2}{3}F_{\mu\rho}F_\nu{}^\rho
          + 32 A_\mu A_\nu \\
       &=- 12g_{\mu\nu} +
           \tfrac{2}{3}\left( F_{\mu\rho}F_\nu{}^\rho
              -\tfrac{1}{8}g_{\mu\nu}F_{\rho\sigma}F^{\rho\sigma}
              \right)
           + 32 \left(A_\mu A_\nu-g_{\mu\nu}A_\rho A^\rho\right)
\end{aligned}
\end{equation}
where we have used \reef{consmav} to get the last line.
These equations of motion come from the Lagrangian
\be
S=\int d^4 x\sqrt{-g}\left[
   R-\tfrac{1}{3}F_{\mu\nu}F^{\mu\nu}
   -32A_\mu A^\mu +24 \right]
\ee
which describes a metric coupled to a massive vector field with
$m^2=48$, provided that we impose the conditions \reef{consmav} by
hand. We will return to this truncation to construct
solutions of $D=11$ supergravity in the next section.

\section{Solutions}

As an application we construct solutions of $D=11$ supergravity by
constructing solutions to the four-dimensional equations given in
\reef{consmav} and \reef{eommav}. We consider the ansatz given by
\begin{equation}
\begin{aligned}\label{jkl}
   ds^2 &= -\alpha^2\rho^{2k}(dx^+)^2
       +\frac{d\rho^2}{4\rho^2}
       +\frac{\rho^2}{4}\left(-dx^+dx^-+dx^2\right) \\
   A_1 &= c\rho^k dx^+
\end{aligned}
\end{equation}
We find that $k=3$ with $c^2=\alpha^2$ solves all the equations as
does $k=-4$ with $c^2=15\alpha^2/8$. We can now uplift these solutions
to $D=11$ by setting $U=V=h=\chi=0$, $f=6$, $H_2=\frac{1}{3}*F$ and
$H_3=8*A_1$ and substituting into \reef{KKmetT} and
\reef{KKG4T}. Writing this out explicitly for
the $k=3$ case
case we obtain
\begin{equation}
\label{one}
\begin{aligned}
   ds^2 &= -\alpha^2\rho^{6}(dx^+)^2
           + \frac{d\rho^2}{4\rho^2}
           + \frac{\rho^2}{4}\left(-dx^+dx^-+dx^2\right)
           \\ & \qquad \qquad
           + ds^2(KE_6) + (\eta+\alpha \rho^3dx^+)^2
      \\
   G &= \tfrac{3}{16}\rho^2dx^+\wedge dx^-\wedge d\rho\wedge dx
       +\tfrac{1}{2}\alpha\, dx^+\wedge dx\wedge d(\rho^4\eta)
\end{aligned}
\end{equation}
This solution is in close analogy to the solutions considered
in~\cite{Maldacena:2008wh} and has a non-relativistic conformal
symmetry with dynamical exponent $z=3$
i.e. is invariant under Galilean transformations generated by time and spatial
translations, Galilean boosts, a central mass operator, and scale 
transformations\footnote{Recall that only for dynamical exponent $z=2$ can the algebra be 
enlarged to include an additional special conformal generator. Also note that the $k=-4$ 
solution has a
non-relativistic conformal symmetry with dynamical exponent $z=-4$
and is singular. We note analogous singular solutions also
exist for the theory considered in \cite{Maldacena:2008wh}.}. 
This solution is supersymmetric, generically preserving two
supersymmetries, as explained in \cite{Donos:2009en}.

\section{Skew-Whiffing}

Recall that for each $AdS_4\times M_7$ Freund--Rubin solution there is
another ``skew-whiffed'' solution~\cite{Duff:1984sv}
which can be obtained by reversing the sign of the flux (or
equivalently changing the orientation of $M_7$). With the exception of
the special case where $M_7$ is the round $S^7$, at most only one of
the two solutions is supersymmetric. For example, if we reverse the
sign of the flux in the
supersymmetric $AdS_4\times SE_7$ solution~\reef{sesol} we obtain
another $AdS_4\times SE_7$ solution of $D=11$
supergravity given by
\bea
\label{skew}
  ds^2&=&\tfrac{1}{4}ds^2(AdS_4)+ds^2(SE_7)\nn
  G_4&=&-\tfrac{3}{8}\vol(AdS_4)
\eea
which does not preserve any supersymmetry (provided $SE_7$ is not $S^7$).

By a very small modification of the truncation discussed in
section~\ref{consistentKK} above, we can obtain a second consistent
truncation on $SE_7$ to a $D=4$ theory that contains the
skew-whiffed solution. In particular, we solve
\reef{mancity} by now setting
\be f=6 e^{-6U-V}(-1+h^2+|\chi|^2) \ee
where we have changed the sign of the constant factor (when $U=V=h=\chi=0$).
The rest of the analysis essentially goes through unchanged but the sign propagates
into the $D=4$ action in two places. In \reef{lageinfull}
$\left( 1+h^2 +|\chi|^2 \right)^2\to\left(-1+h^2 +|\chi|^2 \right)^2$ and
$6A_1\wedge H_3\to -6A_1\wedge H_3$. The $AdS_4$ vacuum solution of this theory
now uplifts to the non-supersymmetric skew-whiffed solution. The mass spectrum for this
vacuum can be easily calculated and the only difference from
section~2.2 is that now $m^2_\chi=-8$ and $m^2_h=-8$ corresponding to
operators with $\Delta_\pm=1,2$. As expected this bosonic mass
spectrum is inconsistent with a vacuum preserving $N=2$ $D=4$
supersymmetry since it does not match the bosonic $Osp(2|4)$
multiplet structure.

Despite the fact that the skew-whiffed vacuum is not supersymmetric the
$D=4$ action has the bosonic content consistent with $N=2$
supersymmetry. The analysis of section 2.3 goes through essentially
unchanged, but the sign change in the $D=4$ action, $6A_1\wedge H_3\to
-6A_1\wedge H_3$, means that the gauging is now along Killing vectors
given by
\begin{equation}
\label{ninsw}
\begin{aligned}
  k_0 &=-6\partial_a+4i(\chi\partial_\chi-\bar\chi\partial_{\bar\chi})
  =-24\partial_\sigma-4i(\xi\partial_\xi-\bar\xi\partial_{\bar\xi})\ , \\
  k_1 &= 6\partial_a = 24\partial_\sigma\ .
\end{aligned}
\end{equation}
The corresponding Killing prepotentials $P_I$ are then
\begin{equation}
  P_0 = -24P_\sigma-4P_\xi \ , \qquad
  P_1 = 24P_\sigma \ .
\end{equation}
Substituting these expressions into the general form~\eqref{N=2V} for
the potential $V$ we reproduce~\eqref{Vdef} after the change $\left(
   1+h^2 +|\chi|^2 \right)^2\to\left(-1+h^2 +|\chi|^2
\right)^2$.
For a general $SE_7$ the $AdS_4$ vacuum spontaneously breaks the
$N=2$ supersymmetry\footnote{Here we are assuming
that the truncation at the level of the bosonic fields can be extended
to include the fermions.} of the action with $f=-6$.

As we have already noted, for the special case that $SE_7$ is the
round $S^7$, the corresponding $AdS_4\times S^7$ solutions are
supersymmetric for either sign of the flux. It is interesting to
observe that while the $AdS_4$ vacuum of the truncated theory with
$f=6$ contains modes that fall into $OSp(2|4)$ multiplets, this
is not the case for the $AdS_4$ vacuum of the theory with $f=-6$,
despite the fact that the uplifted solution is (maximally)
supersymmetric.
In particular, while the $f=6$ theory retains an $N=2$ breathing mode
multiplet together with the supergravity multiplet, in the $f=-6$
theory the modes corresponding to the $h$ and $\chi$ fields are no
longer part of the breathing multiplet but instead are part of the
$N=8$ graviton supermultiplet.  Nonetheless this leads to a consistent
truncation.
This is a novel and interesting phenomenon that would be worth
investigating further, including from the dual SCFT point of view.

Many of the additional truncations of the $N=2$ theory that we considered in
section~\ref{other-trunc} have similar analogues in the
skew-whiffed theory with only some minor obvious sign changes
required. For example, the $D=4$ action that contains the
non-supersymmetric skew-whiffed weak $G_2$ case can be written in a
manifestly $N=1$ language and we find that the only difference is that
the superpotential $W = 4\sqrt{2}(7\phi^2 -3)\to 4\sqrt{2}(7\phi^2
+3)$. For the reduction to the massive vector field, we should now set
$U=V=h=\chi=0$, $f=-6$, $H_2=-\frac{1}{3}*F_2$ and $H_3=-8*A_1$. These
sign changes mean that when we uplift the solution \reef{jkl} to
$D=11$ we obtain the solution \reef{one} but with the sign of the
four-form flux reversed. Note however, as noticed in 
\cite{Denef:2009tp}, it is no longer possible to truncate to the field content 
of minimal gauged supergravity as in section 3.1.

Recently KK reductions of $AdS_4\times SE_7$ solutions were considered
at the linearised level~\cite{Denef:2009tp} and it was shown that, for
the skew-whiffed solution, the modes corresponding to the massless
gauge-field ${\cal A}_1$ and the complex scalar $\chi$ lead to a $D=4$
theory that exhibits holographic superconductivity. Indeed,
at the linearised level, in our analysis we can set $U=V=h=H_3=0$ with
$F_2=\pm *H_2$ where the upper (lower) sign corresponds to the
supersymmetric (skew-whiffed) truncation. Writing $A_1={\cal
  A}/2$, $\chi=\sqrt{2/3}\phi$, the linearised action is given by
\begin{equation}
\begin{aligned}
  S &= \int d^4 x\sqrt{-g}\Big[
      R + 24 -\frac{1}{4}{\cal F}_{\mu\nu}{\cal F}^{\mu\nu}
      - |D\phi|^2-m^2|\phi|^2\Big]
\end{aligned}
\end{equation}
with $D\phi=d\phi-2i{\cal A} \phi$ 
and $m^2=40,-8$ for the supersymmetric and skew-whiffed case,
respectively. This is in agreement with
\cite{Denef:2009tp}, upon setting $M^2=2$, $L^2=1/4$, $q=2$ and $g=1$ in their
equation (1). In particular, for the skew-whiffed solution, there are
solutions of this linearised theory corresponding to holographic
superconductors. Our generalised, non-linear and consistently
truncated action for the skew-whiffed solutions thus provides an ideal
set up to extend the work of \cite{Denef:2009tp} to obtain analogous
exact solutions of $D=11$ supergravity.

\section{Discussion}

In this paper we have considered consistent truncations on
Freund--Rubin backgrounds, keeping the breathing mode and with varying
degrees of supersymmetry. We have shown that for $AdS_4\times M_7$
solutions of $D=11$ supergravity where $M_7$ is an Einstein space, it
is always consistent to truncate the KK spectrum to the graviton plus
the breathing mode, which is dual to an operator in the dual CFT with
$\Delta=6$. For $AdS_4\times M_7$ solutions with $N=1$ and $N=2$
supersymmetry, where $M_7$ has weak $G_2$ holonomy or is a
Sasaki--Einstein seven manifold, respectively, we have also shown that
it is consistent to truncate to the massless graviton supermultiplet
combined with the supermultiplet containing the massive breathing
mode. In both cases, the KK ansatz contains the constant KK modes
associated with the weak $G_2$ or the Sasaki--Einstein structure.

Moving to $AdS_4\times M_7$ solutions with $N=3$ supersymmetry, where
$M_7$ is tri-Sasakian, it is natural to expect that a similar story
unfolds. Recall that a tri-Sasakian manifold has an $SO(3)$ group of
isometries corresponding to $SO(3)$ $R$-symmetry. By writing down a KK
ansatz that incorporates the constant modes associated with the
tri-Sasaki structure we strongly suspect that it will be possible to
obtain a consistent KK truncation with $N=3$ supersymmetry. Such a
truncation would retain the fields of the massless graviton
supermultiplet (table 3 of \cite{Fre':1999xp}) which consist of the
graviton and the $SO(3)$ vector fields,
and the breathing mode supermultiplet, which now sits in a long
gravitino multiplet (table 2 of \cite{Fre':1999xp} with $J_0=0$)
consisting of six massive vectors, transforming in two spin-one
representations of $SO(3)$, four scalars in the spin-zero
representation, and ten scalars transforming in two spin-two
representations.

Following this pattern one is led to consider the maximally
supersymmetric $AdS_4\times S^7$ solution with $N=8$ supersymmetry. It
is again natural to conjecture that there is an
analogous consistent KK truncation that extends the one
containing just the $N=8$ graviton supermultiplet \cite{de Wit:1986iy},
i.e. $N=8$ $SO(8)$ gauged supergravity, to also include the $N=8$
supermultiplet containing the breathing mode.
Using the results of
\cite{Casher:1984ym}
or \cite{Duff:1986hr} we conclude that the bosonic fields of this
supermultiplet consist of scalars in the ${\bf 294_v}$, ${\bf 840_s'}$, {\bf
300}, ${\bf 35_s}$ and {\bf 1} irreps of $SO(8)$, where the singlet is
the breathing mode, vectors in the ${\bf 567_v}$, {\bf 350} and {\bf 28}
irreps and massive spin-two fields in the ${\bf 35_v}$ irrep.
A particularly interesting feature is the appearance of massive
spin-two fields in addition to the graviton. This is remarkable since
some general arguments have been put forward, for instance
in~\cite{Duff:1989ea}, that it is not possible to have consistent
theories of a  finite number of massive and massless spin-two fields.
However, for instance, the group theory arguments
in~\cite{Duff:1989ea}, as for conventional $N=8$ $SO(8)$ supergravity,
are not directly applicable here, and furthermore we are led to a
theory with a very particular matter content, which suggests a
picture where consistency arises from particular conspiracies among
the fields, and perhaps depending crucially on the existence of an
$AdS$ vacuum.
If this putative theory exists, it may also not be possible
to further truncate the theory while keeping massive spin-two
fields. It is worth pointing out that unlike the cases we have
studied in this paper, and the tri-Sasakian case mentioned above, it
is much less clear how to directly construct the KK truncation ansatz
for this case.

Let us now return to the $AdS_5\times M_5$ solutions of type IIB
supergravity where $M_5$ is Einstein. Once again there is a
consistent KK truncation that keeps the graviton and the breathing
mode which is now dual to an operator with $\Delta=8$. If $M_5$ is
Sasaki--Einstein then it is possible to
generalise the ans\"atze of \cite{Buchel:2006gb} and of
\cite{Maldacena:2008wh} to obtain a consistent KK truncation
that includes the bosonic fields of the $N=1$ graviton multiplet plus the
breathing mode multiplet. We will report on the details of this
in \cite{gkvw}.

For the special case when $M_5=S^5$ we are led to conjecture that
there is a consistent truncation to the massless graviton
supermultiplet, i.e. the fields of maximal $SO(6)$ gauged
supergravity, combined with the massive breathing mode multiplet whose
field content can be obtained from \cite{Kim:1985ez}: the bosonic
fields consist of scalars in the ${\bf 105}$, ${\bf 126_C}$, ${\bf
  20_C}$, ${\bf 84}$, ${\bf 10_C}$, ${\bf 1}$ irreps of $SU(4)$, where
the breathing mode is again the singlet, vectors in the ${\bf 175}$,
${\bf 64_C}$, ${\bf 15}$ irreps, two-forms in  the ${\bf 6_C}$, ${\bf
  {45_C}}$, ${\bf 50_C}$ irreps and massive spin-two fields in the
${\bf 20}$ irrep. Note that for this case the operator dual to the
breathing mode has been argued to be dual to an operator in $N=4$
super Yang-Mills theory of the form $Tr F^4+\dots$, where here $F$ is
the $N=4$ Yang-Mills field strength, and it has been argued that its
detailed form can be obtained from expanding the Dirac--Born--Infeld
action for the D3-brane
\cite{Gubser:1998kv}\cite{Intriligator:1999ai}\cite{Danielsson:2000ze}.

In a similar spirit we can consider
$AdS_7\times S^4$ solutions of $D=11$ supergravity. There is a known
consistent truncation \cite{Bremer:1998zp} that keeps the graviton
and the breathing mode which is now dual to an operator with scaling
dimension $\Delta=12$. If this can be extended to include the full
$N=8$ supermultiplets then there would be a consistent KK truncation
extending the known one to maximal $SO(5)$ gauged supergravity
\cite{Nastase:1999cb}\cite{Nastase:1999kf} to also include the
breathing mode supermultiplet. The field content of this latter multiplet
can be found in \cite{Leigh:1998kt} (based on the results of
\cite{vanNieuwenhuizen:1984iz} \cite{Gunaydin:1984wc}): we find
scalars in the {\bf 55}, {\bf 35} and {\bf 1} irreps of $SO(5)$,
where the singlet is the breathing mode, vectors in the {\bf 81}
and {\bf 10} irreps, three-forms satisfying self-dual equations in the {\bf
30} and {\bf 5} irreps, two-forms in the {\bf 35} irrep and massive
spin-two fields in the {\bf 14} irrep.

It would also be interesting to see if similar results can be obtained for classes of supersymmetric
$AdS$ solutions outside of the Freund-Rubin class that we have been considering so far.
The KK truncations to the massless graviton supermultiplets
for the class of $N=2$ and $N=1$ $AdS_5$ solutions
of $D=11$ supergravity classified in \cite{Lin:2004nb}
and \cite{Gauntlett:2004zh}
were presented in \cite{Gauntlett:2007sm} and \cite{Gauntlett:2006ai}, respectively.
Similarly, the KK truncations for the class of $N=2$ $AdS_4$ solutions of $D=11$ supergravity
and the class of $N=1$ $AdS_5$ solutions of type IIB which were classified in \cite{Gauntlett:2006ux}
and \cite{Gauntlett:2005ww},
respectively, were presented in \cite{gv}. It would be interesting to extend these KK truncations to
also include breathing mode multiplets.

Finally, it would be desirable to have an argument from the SCFT side
of the correspondence as to why the KK truncations containing both the
graviton multiplets and the massive breathing mode multiplets are
consistent.

\subsection*{Acknowledgements}
We would like to thank Frederik Denef, Mike Duff, Ami Hanany, Sean Hartnoll, Chris
Hull, Mukund Rangamani, James Sparks, Kelly Stelle, Arkady Tseytlin
and Toby Wiseman for helpful discussions. JPG is supported by an EPSRC
Senior Fellowship and a Royal Society Wolfson Award. OV is supported
by a Spanish Government's MEC-FECYT postdoctoral fellowship, and
partially through MEC grant FIS2008-1980.

\appendix

\section{Supersymmetry of the $AdS_4\times SE_7$ Solution}
\label{app:susy}

In this appendix we show that the solution given by
\begin{equation}
\begin{aligned}
   ds^2 &= \tfrac{1}{4}ds^2(AdS_4)+ds^2(SE_7) \ , \\
   G &= 6\vol_4=\tfrac{3}{8}\vol(AdS_4) \ ,
\end{aligned}
\end{equation}
is supersymmetric given our set of conventions. These are, for
$D=11$ supergravity, the conventions given in \cite{Gauntlett:2002fz},
the structure on $SE_7$ is defined by the forms $\eta$, $J$ and
$\Omega$ satisfying
\begin{equation}
\label{SEstructrued}
\begin{aligned}
   d\eta &= 2J \ , \\
   d \Omega &= 4i\eta\wedge \Omega \ , \\
   \vol(SE_7) &= \eta\wedge\tfrac{1}{3!}J^3
      = \eta\wedge \tfrac{i}{8}\Omega\wedge\Omega^* \ ,
\end{aligned}
\end{equation}
and the $D=11$ volume form is $\epsilon=\vol_4\wedge\vol(SE_7)$.

It will be sufficient to focus on the Poincar\'e supersymmetries. To
do so, we start by rewriting the solution in terms of a Calabi--Yau
fourfold cone metric. We introduce coordinates for the $AdS_4$ space
\bea
\tfrac{1}{4}ds^2(AdS_4)
   = \frac{1}{4}\left(
        \frac{d\rho^2}{\rho^2}
        +\rho^2\eta_{\mu\nu} d\bar \xi^\mu d\bar \xi^\nu \right)
   = \frac{dr^2}{r^2}+r^4 \eta_{\mu\nu}d \xi^\mu d \xi^\nu
\eea
with $\rho=r^2$, $\bar\xi^\mu=2\xi^\mu$ and $\mu=0,1,2$, and define
the four-dimensional volume form $\vol_4=r^5d\xi^0\wedge d\xi^1\wedge
d\xi^2\wedge dr$. The $D=11$ solution can then be recast in the form
\begin{equation}
\label{CYform}
\begin{aligned}
   ds^2 &= H^{-2/3}\eta_{\mu\nu}d\xi^\mu d\xi^\nu
      + H^{1/3}ds^2(C_8) \ , \\
   G &= d\xi^0\wedge d\xi^1\wedge d\xi^2\wedge d(H^{-1})
\end{aligned}
\end{equation}
where we have introduced the cone metric over the $SE_7$ space,
\be
ds^2(C_8)= dr^2+r^2ds^2(SE_7)~,
\ee
and $H=r^{-6}$ is harmonic on $C_8$. The eleven-dimensional volume form
is then $\epsilon = H^{1/3}d\xi^0\wedge d\xi^1\wedge
d\xi^2\wedge\vol(C_8)$ where $\vol(C_8)=r^7dr\wedge\vol(SE_7)$.

The Sasaki--Einstein structure~\eqref{SEstructrued} defines a unique
Calabi--Yau structure on the cone given by the $SU(4)$ invariant
tensors
\begin{equation}
\label{CCY}
\begin{aligned}
   J_{CY} &= rdr\wedge\eta+r^2J \ , \\
   \Omega_{CY} &= r^3(dr+ir\eta)\wedge\Omega \ ,
\end{aligned}
\end{equation}
determined by requiring the closure of $J_{CY}$ and $\Omega_{CY}$ to
be equivalent to $d\eta =2J$ and $d\Omega=4i\eta\wedge\Omega$. In
particular, we then find
\begin{equation}
\label{vol-rel}
   \vol(C_8) = \tfrac{1}{4!}J_{CY}^4
      = \tfrac{1}{16}\Omega_{CY}\wedge\Omega_{CY}^* \ .
\end{equation}

We now turn to the supersymmetry. We introduce a $D=11$ orthonormal
frame:
\be
   e^\mu=H^{-1/3}d\xi^\mu \ , \qquad
   e^{a+2} =H^{1/6}g^a \ , \quad a=1,\dots,8 \ ,
\ee
where $g^a$ is an orthonormal frame for the cone metric. Following the
conventions of~\cite{Gauntlett:2002fz}, by definition
$\epsilon=e^0\wedge e^1\wedge\dots\wedge e^{10}$ and so
\begin{equation}
\label{Cvol}
   \vol(C_8) = g^1\wedge g^2 \wedge \dots \wedge g^8 \ .
\end{equation}
We can then decompose the $D=11$ gamma-matrices as
\begin{equation}
\begin{aligned}
   \Gamma_\mu &= \tau_\mu\otimes \gamma_{(8)} \ , && \\
   \Gamma_{a+2} &= \mathbf{1}\otimes \gamma_a \ ,
      & \qquad a &=1,\dots, 8
\end{aligned}
\end{equation}
with $\tau_{012}=1$ and where $\gamma_{(8)}=\gamma_1\gamma_2\dots
\gamma_8$ is the chirality operator in $D=8$. The $D=11$ supersymmetry
equations given in~\cite{Gauntlett:2002fz} are satisfied by a solution
of the form~\eqref{CYform} provided the supersymmetry transformation
parameter satisfies the gamma-matrix projection
condition
\be
\Gamma_{012}\epsilon=\epsilon
\quad\Leftrightarrow\quad
\Gamma_{34\dots 10}\epsilon=\epsilon~.
\ee
More precisely, there are Poincar\'e Killing spinors of the form
\be
\epsilon = H^{-1/6}\alpha\otimes \beta \ ,
\ee
where $\alpha$ is a constant two-component Majorana spinor in $D=3$
and $\beta$ is a 16-component Majorana--Weyl spinor in $D=8$ satisfying
\begin{equation}
   \nabla_a\beta = 0 \ , \qquad
   \gamma_{(8)}\beta = \beta \ .
\end{equation}
For there to be two independent solutions $\beta_{(i)}$ with $i=1,2$,
the cone metric must be Calabi--Yau. In particular, the $\beta_{(i)}$ can
be chosen to be orthogonal and the Calabi--Yau structure $J_{CY}$ and
$\Omega_{CY}$ can be written as bilinears in
$\beta_{(i)}$. Specifically, one can choose a frame $\{g^a\}$ and
spinor projections exactly as in appendix~B of~\cite{Gauntlett:2003cy}
such that
\begin{equation}
\begin{aligned}
   J_{CY} &= g^{12}+g^{34}+g^{56}+g^{78} \ , \\
   \Omega_{CY}&= (g^1+ig^2)\wedge(g^3+ig^4)\wedge(g^5+ig^6)
       \wedge(g^7+ig^8) \ .
\end{aligned}
\end{equation}
Crucially, from~\eqref{Cvol}, we see these satisfy the orientation
relation~\eqref{vol-rel}. Thus the Calabi--Yau structure~\eqref{CCY}
on the cone $C_8$ defined by the Sasaki--Einstein
structure~\eqref{SEstructrued} is indeed of the type required for the
solution to be supersymmetric.

Note that if one takes the skew-whiffed solution where
$G_4=-\frac{3}{8}\vol(AdS_4)$, supersymmetry would then imply
$\gamma_{(8)}\beta=-\beta$. This would in turn require a
Calabi--Yau structure $(J'_{CY},\Omega'_{CY})$ on $C_8$ satisfying
$\vol(C_8)=-\frac{1}{4!}J^{\prime 4}_{CY}=
-\frac{1}{16}\Omega'\wedge\Omega^{\prime\star}$. The structure defined
by the Sasaki--Einstein manifold is not of this type, and hence the
skew-whiffed solution is generically not supersymmetric.

\section{Details on the KK reduction}
\label{app:KK}

As discussed in the main text, our ansatz for the metric of $D=11$
supergravity is given by
\be\label{KKmet}
ds^2=ds^2_4+e^{2U}ds^2(KE_6)+e^{2V}(\eta+A_1)\otimes(\eta +A_1)
\ee
while for the four-form we consider
\begin{equation}
\begin{aligned}
   G_4 &= f \textrm{vol}_4 +H_3 \wedge(\eta+A_1) + H_2 \wedge J
          + H_1 \wedge J \wedge (\eta+A_1) + 2h J \wedge J
          \\ & \qquad \qquad \qquad
          +{\sqrt 3}\left[ \chi_1\wedge\Omega
                + \chi(\eta+A_1)\wedge\Omega+ \textrm{c.c.} \right] \; .
\end{aligned}
\end{equation}
For the $D=11$ volume-form we choose $\epsilon=e^{6U+V}\vol_4\wedge
\vol(KE_6)\wedge \eta$, where $\vol_4$ is the $D=4$ volume form. In both $D=11$ and $D=4$ we use
a mostly plus signature convention.

We now substitute this ansatz into the
equations of motion of $D=11$ supergravity. The Bianchi identity
$dG_4=0$ is satisfied provided
\begin{align}
\label{Bianchi1} dH_3 &= 0 \ , \\
\label{Bianchi2}dH_2 &= 2H_3+H_1 \wedge F_2 \ , \\
\label{Bianchi4} H_1 &= dh \ , \\
\label{chi1} \chi_1 &= -\tfrac{i}{4}D\chi \ ,
\end{align}
where $F_2\equiv dA_1$, $D\chi\equiv d\chi-4iA_1\chi$ and we note
that \reef{Bianchi1} follows from \reef{Bianchi2} and
\reef{Bianchi4}.
Note that using \reef{Bianchi4} and \reef{chi1} we can write the
four-form as
\begin{equation}
\label{KKG4}
\begin{aligned}
   G_4 &= f \vol_4 +H_3 \wedge(\eta+A_1) + H_2 \wedge J
        + dh \wedge J \wedge (\eta+A_1) + 2h J \wedge J
        \\ & \qquad \qquad \qquad
        + {\sqrt 3}\left[
           \chi(\eta+A_1)\wedge\Omega-\tfrac{i}{4}D\chi\wedge\Omega
           + \textrm{c.c.} \right] \; .
\end{aligned}
\end{equation}
We solve equations \reef{Bianchi1} and \reef{Bianchi2}
by introducing potentials $B_2$ and $B_1$ via
\begin{equation}
\begin{aligned}
H_3 &=dB_2 \ , \\
H_2 &=dB_1+2B_2+hF_2 \ .
\end{aligned}
\end{equation}
Similarly the equation of motion for the four-form, $d*_{11} G_4
+\tfrac{1}{2}G_4 \wedge G_4=0$, is also satisfied if
\begin{align}
\label{geom1}
   & d \left( e^{6U-V} *H_3 \right) -e^{6U+V}f F_2 +
       6e^{2U+V} *H_2 + 12hH_2 +\frac{3i}{2}D\chi\wedge D\chi^\ast =0 \\
   & d\left( e^{2U+V} *H_2 \right) + 2dh\wedge H_2 + 4hH_3 =0\label{manu} \\
   & d\left( e^{2U-V} *dh \right) +e^{2U+V} *H_2 \wedge F_2 +H_2
      \wedge H_2 + 4h \left( f + 4e^{-2U +V} \right) \textrm{vol}_4 =0 \\
   & d[e^{6U+V}f-6(h^2+|\chi|^2)] =0 \label{mancity}\\
   & D\left(e^V\ast D\chi\right)+iH_3\wedge D\chi+
      4\chi(f+4e^{-V}){\rm vol}_4 =0\label{hull} \ .
\end{align}
One can show that \reef{manu} can be obtained by acting with $d$
on \reef{geom1}. We can solve \reef{mancity} by setting
\be f=6 e^{-6U-V}(\pm 1+h^2+|\chi|^2) \ee where the constant factor of
$\pm6$ (when $U=V=h=\chi=0$) is chosen as a convenient normalisation.
The upper sign corresponds to reducing to a $D=4$ theory that contains
the supesymmetric $AdS_4\times SE_7$ solution of $D=11$ supergravity while
the lower sign corresponds to the skew-whiffed $AdS_4\times SE_7$ solution,
which generically doesn't preserve any supersymmetry.

Finally we consider the $D=11$ Einstein equations:
\be \label{rhsEinstein}
    R_{AB}=
       \tfrac{1}{12}G_{4\,AC_1C_2C_3}G_{4\,B}{}^{C_1C_2C_3}
       -\tfrac{1}{144}g_{AB}G_{4\,C_1C_2C_3C_4}G_4^{C_1C_2C_3C_4} \ .
\ee
To calculate the Ricci tensor for the $D=11$ metric we use the orthonormal frame
\begin{equation}
\label{KKframe}
\begin{aligned}
   \bar e^\alpha &= e^\alpha \; , & \alpha &= 0,1,2,3 \ , \\
   \bar e^i &= e^{U}e^i \; , & i &= 1,\ldots,6 \ , \\
   \bar e^7 &= e^V \hat e^7 \equiv e^V ( \eta + A_1) \ .
\end{aligned}
\end{equation}
We then observe that the corresponding spin
connection can be written
\begin{equation}
\begin{aligned}
   \bar \omega^{\alpha \beta} &= \omega^{\alpha \beta} -\tfrac{1}{2}e^{2V}F^{\alpha \beta} \hat e^7 \\
   \bar \omega^{\alpha i} &= -e^U \partial^\alpha U e^i \\
   \bar \omega^{\alpha 7} &= -e^V \partial^\alpha V \hat  e^7 -\tfrac{1}{2}e^{V}F^{\alpha}{}_{ \beta} e^\beta \\
   \bar \omega^{ij} &= \omega^{ij} -e^{2V-2U} J^{ij} \hat e^7 \\
   \bar \omega^{i7} &= -e^{V-U} J^i{}_je^j
\end{aligned}
\end{equation}
After some computation we find that the components of the Ricci
tensor, $\bar R_{AB}$, are given by
\begin{equation}
\begin{aligned}
   \bar R_{\alpha \beta} &= R_{\alpha \beta} -6 \left(\nabla_\beta
      \nabla_\alpha U + \partial_\alpha U \partial_\beta U \right)
   -\left(\nabla_\beta \nabla_\alpha V + \partial_\alpha
      V \partial_\beta V \right) -\tfrac{1}{2}e^{2V} F_{\alpha \gamma}
   F_\beta{}^\gamma \\
   \bar R_{\alpha i} &= 0 \\
   \bar R_{\alpha 7} &= -\tfrac{1}{2} e^{-2V-6U} \nabla_\gamma
   \left(e^{3V+6U} F^{\gamma \alpha} \right) \\
   \bar R_{ij} &= \delta_{ij} \left[8e^{-2U} -2e^{2V-4U}
      -\nabla_\gamma \nabla^\gamma U -6\partial_\gamma
      U \partial^\gamma U -\partial_\gamma U \partial^\gamma V \right]
   \\
   \bar R_{i7} &= 0 \\
   \bar R_{77} &= 6e^{2V-4U} -\nabla_\gamma \nabla^\gamma V
   -6\partial_\gamma U \partial^\gamma V -\partial_\gamma
   V \partial^\gamma V +\tfrac{1}{4}e^{2V} F_{\alpha \beta}
   F^{\alpha\beta}
\end{aligned}
\end{equation}

Using these results we find that the $D=11$ Einstein equations
\reef{rhsEinstein} reduce to the following four equations in $D=4$:
\begin{equation}
\label{Ein1chi}
\begin{aligned}
   R_{\alpha \beta} &= 6 \left(\nabla_\beta \nabla_\alpha U
      + \partial_\alpha U \partial_\beta U \right) +\left(\nabla_\beta
      \nabla_\alpha V + \partial_\alpha V \partial_\beta V \right)
   \\ & \qquad
   + \tfrac{3}{2} e^{-4U-2V}  \left( \nabla_{\alpha}h \nabla_{\beta}h
      -\tfrac{1}{3} \eta_{\alpha \beta} \nabla_{\lambda}h
      \nabla^{\lambda}h \right)
   \\ & \qquad
   + \tfrac{3}{4} e^{-6U} \left[ (D_\alpha \chi)(D_\beta \chi^*)
      +(D_\beta \chi)(D_\alpha \chi^*) -\tfrac{2}{3} \eta_{\alpha
        \beta}(D_\gamma \chi)(D^\gamma \chi^*) \right]
   \\ & \qquad
   -2 \eta_{\alpha \beta} \left(e^{-8U}4h^2 +\tfrac{1}{6} f^2
      +4e^{-6U-2V}|\chi|^2 \right) +\tfrac{1}{2}e^{2V} F_{\alpha
     \gamma} F_\beta{}^\gamma
   \\ & \qquad
   + \tfrac{1}{4} e^{-2V} \left( H_{\alpha \lambda \mu} H_{\beta}{}^{
        \lambda \mu} -\tfrac{1}{9} \eta_{\alpha \beta} H_{\lambda \mu
        \nu} H^{\lambda \mu \nu} \right)
   \\ & \qquad
   + \tfrac{3}{2} e^{-4U} \left( H_{\alpha \lambda}
      H_{\beta}{}^{\lambda} -\tfrac{1}{6} \eta_{\alpha \beta}
      H_{\lambda \mu} H^{\lambda \mu} \right)
\end{aligned}
\end{equation}
\begin{equation}
\label{Ein2chi}
\begin{aligned}
   \nabla_\gamma &\left(e^{3V+6U} F^\gamma{}_\alpha \right) =
   \tfrac{1}{6} e^{6U+V}f \epsilon_{\alpha\beta\gamma\delta}
   H^{\beta\gamma\delta} +3e^{2U+V} H_{\alpha\beta} \nabla^\beta h
   \qquad \qquad \qquad \\ & \qquad \qquad \qquad \qquad
   + 6i e^V \left[ \chi^* D_\alpha \chi - \chi D_\alpha \chi^* \right]
\end{aligned}
\end{equation}
\begin{equation}
\label{Ein4}
\begin{aligned}
   \nabla_\gamma \nabla^\gamma U
       +{}& 6\partial_\gamma U \partial^\gamma U
       + \partial_\gamma U \partial^\gamma V
       + \tfrac{1}{4}e^{-6U} (D_\gamma \chi)(D^\gamma\chi^*)
       - \tfrac{1}{36}e^{-2V} H_{\alpha\beta\gamma}H^{\alpha\beta\gamma}
       \\ &
       - 8e^{-2U} +2e^{2V-4U}+8e^{-8U}h^2
       + \tfrac{1}{6}f^2 +{4} e^{-6U-2V} |\chi|^2 = 0
\end{aligned}
\end{equation}
\begin{equation}
\label{Ein4chi}
\begin{aligned}
   \nabla_\gamma \nabla^\gamma V +{}& 6\partial_\gamma U \partial^\gamma V
       + \partial_\gamma V\partial^\gamma V
       + e^{-4U-2V} \nabla_{\lambda}h \nabla^{\lambda}h
       - \tfrac{1}{2}e^{-6U} (D_\gamma \chi)(D^\gamma \chi^*)
       \\ &
       - 6e^{2V-4U}  -8e^{-8U}h^2 + \tfrac{1}{6}f^2
       + {16} e^{-6U-2V} |\chi|^2
       \\ &
       - \tfrac{1}{4}e^{2V} F_{\alpha \beta}F^{\alpha\beta}
       + \tfrac{1}{18} e^{-2V} H_{\alpha\beta\gamma}H^{\alpha\beta\gamma}
       -\tfrac{1}{4} e^{-4U} H_{\alpha\beta} H^{\alpha\beta}
       = 0
\end{aligned}
\end{equation}

All of the dependence on the internal $SE_7$ space has
dropped out. In particular any solution to the $D=4$ field equations
(\ref{Bianchi1})--(\ref{chi1}), (\ref{geom1})--\reef{hull},
(\ref{Ein1chi})--(\ref{Ein4chi}) gives rise to an exact solution to
the equations of motion of $D=11$ supergravity. Thus the KK ansatz
(\ref{KKmet}), (\ref{KKG4}) is consistent.


\section{$N=2$ supergravity}
\label{N2susy}

The bosonic part of the general gauged $\mathcal{N}=2$ supergravity
action coupled to vector and hypermultiplets is given
by~\cite{Andrianopoli:1996cm,Louis:2002ny}
\begin{equation}
\label{N=2action}
\begin{aligned}
   S = \int \tfrac{1}{2}R\ast 1
      +{}& g_{i\bar{j}}Dt^i\wedge\ast D\bar{t}^j
      + h_{uv}Dq^u\wedge\ast Dq^v \\
      & + \tfrac{1}{2}\im\mathcal{N}_{IJ}F^I\wedge\ast F^J
      + \tfrac{1}{2}\re\mathcal{N}_{IJ}F^I\wedge F^J
      - V \ .
\end{aligned}
\end{equation}
Here $t^i$, $i=1,\dots, n_V$ are the complex scalar fields in the $n_V$ vector multiplets
parameterizing a special K\"ahler manifold with metric $g_{i\bar{j}}$,
while $q^u$, $u=1,\dots, 4n_H$, are the real scalar fields in the $n_H$ hypermultiplets
parameterizing a quaternionic manifold with metric $h_{uv}$. The
two-forms $F^I=dA^I$ with $I=0,1,\dots,n_V$ are the gauge field strengths for the
vector multiplet and graviphoton potentials $A^I$. In the gauged theory
\begin{equation}
   D_\mu t^i = \partial_\mu t^i - k_I^i A_\mu^I \ , \qquad
   D_\mu q^u = \partial_\mu q^u - k_I^u A_\mu^I \ , \qquad
\end{equation}
where $k_I^i$ and $k_I^u$ are Killing vectors on the special K\"ahler
and quaternionic manifolds. For the theories appearing in this paper
$k_I^i=0$.

The metric on the special K\"ahler manifold and the gauge kinetic
terms can both be written in terms of a holomorphic
prepotential $\mathcal{F}(X)$ where $X^I(t)$ are homogeneous
coordinates on the manifold and which is a homogeneous function of
degree two. Explicitly the K\"ahler potential and
$\mathcal{N}_{IJ}$ matrix are given by
\begin{equation}
\begin{aligned}
   K_V &= - \log\left(
       i\bar{X}^I\mathcal{F}_I-iX^I\bar{\mathcal{F}}_I\right) \ , \\
  \mathcal{N}_{IJ} &= \bar{\mathcal{F}}_{IJ}
       + 2i\frac{(\im\mathcal{F}_{IK})(\im\mathcal{F}_{JL})X^KX^L}
       {(\im\mathcal{F}_{AB})X^AX^B} \ ,
\end{aligned}
\end{equation}
with $\mathcal{F}_I=\partial_I\mathcal{F}$ and
$\mathcal{F}_{IJ}=\partial_I\partial_J\mathcal{F}$.
Under symplectic transformations acting on the gauge fields $F^I$ and
the generalised duals $G_I=\partial\mathcal{L}/\partial A^I$, where
$\mathcal{L}$ is the scalar Lagrangian for the supergravity
action~\eqref{N=2action}, one has
\begin{equation}
   \begin{pmatrix} F^I \\ G_I \end{pmatrix}
      \mapsto \begin{pmatrix}
          \tilde{F}^I \\ \tilde{G}_I \end{pmatrix}
       = \begin{pmatrix} A & B \\ C & D \end{pmatrix}
          \begin{pmatrix}
             F^I \\ G_I \end{pmatrix} \ , \\
\end{equation}
where $A^TD-C^TB=1$, $A^TC=C^TA$ and $B^TD=D^TB$. The
$(X^I,\mathcal{F}_I)$ coordinates and $\mathcal{N}_{IJ}$ then
transform as
\begin{equation}
\label{eq:symplectic}
\begin{aligned}
   \begin{pmatrix} X^I \\ \mathcal{F}_I \end{pmatrix}
       & \mapsto \begin{pmatrix}
          \tilde{X}^I \\ \tilde{\mathcal{F}}_I \end{pmatrix}
       = \begin{pmatrix} A & B \\ C & D \end{pmatrix}
          \begin{pmatrix}
             X^I \\ \mathcal{F}_I \end{pmatrix} \ , \\
   \mathcal{\mathcal{N}}
       &\mapsto \tilde{\mathcal{N}}
       = (C+D\mathcal{N})(A+B\mathcal{N})^{-1} \ .
\end{aligned}
\end{equation}

The quaternionic manifold has $SU(2)\times Sp(2n_H)$ special holonomy,
so, as in for example \cite{Lukas:1998tt}, one can introduce vielbeins
$V^{A\alpha}$ where $A=1,2$ and $\alpha=1,\dots,n_H$ such that
$h_{uv}=V^{A\alpha}_uV^{B\beta}_v\epsilon_{AB}\bbC_{\alpha\beta}$
where $\epsilon_{12}=-1$ and $\bbC_{\alpha\beta}$ is the constant
symplectic form for $Sp(n_H)$. This defines $SU(2)$ and $Sp(n_V)$
connections via $dV^{Aa}+\omega^A_{\ B}\wedge V^{Ba}+\Delta^a_{\
  b}\wedge V^{Ab}=0$. The triplet of K\"ahler forms can then be
written as
\begin{equation}
   K = K^x \left(-\tfrac{i}{2}\sigma^x\right)
     = -\tfrac{1}{2}\left(d\omega+\omega\wedge\omega\right) \ ,
\end{equation}
where $\sigma^x$ with $x=1,2,3$ are the Pauli matrices. The
corresponding complex structures $(J^x)^u_{\ v}=h^{uw}(K^x)_{wv}$
then satisfy the quaternion algebra.

Given the Killing vectors $k^u_I$ one can then introduce triplets of
Killing prepotentials $P_I=P_I^x\left(-\frac{i}{2}\sigma^x\right)$
satisfying
\begin{equation}
   i_{k_I} K = dP + [ \omega , P ] \ .
\end{equation}
If the gauging is only in the hypermultiplet sector then the
potential $V$ in the action~\eqref{N=2action} is given by
\begin{equation}
\label{N=2V}
   V = e^{K_V} X^I\bar{X}^J \left(4h_{uv}k^u_Ik^v_J\right)
      - \left(\tfrac{1}{2}({\rm Im}\mathcal{N})^{-1IJ}
         +4e^{K_V}X^I\bar{X}^J\right)P^x_IP^x_J \ .
\end{equation}

It is well-known that the universal hypermultiplet parameterizes a
$SU(2,1)/U(2)$ coset. One can identify the particular quaternionic
geometry as follows~\cite{strom}. The metric $h_{uv}$ can be written
as
\begin{equation}
\label{univhyper}
   h_{uv} dq^u dq^v = \frac{1}{4\rho^2} d\rho^2
       + \frac{1}{4\rho^2} \left[
          d\sigma-i(\xi d\bar{\xi}-\bar{\xi} d\xi)\right]^2
       + \frac{1}{\rho} d\xi d\bar{\xi} \; ,
\end{equation}
which has Ricci tensor equal to minus six times the metric.
Introducing the one-forms
\begin{equation}
   \uu = \frac{d\xi}{\sqrt{\rho}}, \qquad
   \vv = \frac{1}{2\rho} \left(
      d\rho + id\s + \xi d\bar{\xi} - \bar{\xi} d\xi \right)
\end{equation}
one can write
\begin{equation}
   V^{A\alpha} = \frac{1}{\sqrt{2}}
       \left( \begin{array}{cc}
          \uu & \ \vb \\ \vv & \ -\ub
       \end{array} \right)^{A\alpha}
\end{equation}
and $h_{uv}=\epsilon_{\alpha\beta}\bbC_{AB}V^{A\alpha}_uV^{B\beta}_v$ with
the constant symplectic form $\bbC$ having components $\bbC_{12}=1$.
We also find
\begin{equation}\label{conn1}
   {\omega^A}_B = {\left( \begin{array}{cc}
          \frac{1}{4}(\vv-\vb) & -\uu \\ \ub & -\frac{1}{4}(\vv-\vb)
        \end{array} \right)^A}_B ,
   \quad
   {\Delta^\alpha}_\beta = {\left( \begin{array}{cc}
          -\frac{3}{4}(\vv-\vb) & 0 \\ 0 & \frac{3}{4}(\vv-\vb)
        \end{array} \right)^\alpha}_\beta \; .
\end{equation}
and
\begin{equation}
   {K^A}_B = {\left( \begin{array}{cc}
           \frac{1}{2}(\uu \wedge \ub - \vv \wedge \vb)  &  \uu \wedge \vb \\
           \vv \wedge \ub  &  - \frac{1}{2}(\uu \wedge \ub - \vv \wedge \vb)
         \end{array} \right)^A}_B \; .
\end{equation}

\end{document}